\documentclass[12pt,preprint]{aastex}
\begin{document}
%
\title{Sky Brightness at Weihai Observatory of Shandong University}

\shorttitle{Sky brightness}
\shortauthors{D. F. Guo et al.}


\author{Di-Fu Guo } \author{Shao-Ming Hu \altaffilmark{*}} \author{Xu Chen}\author{Dong-Yang Gao} \author{Jun-Ju Du}
\affil{Shandong Provincial Key Laboratory of Optical Astronomy and Solar-Terrestrial Environment, Institute of Space Sciences, School of Space Science and Physics, Shandong University, Weihai, 264209, China}
\altaffiltext{*}{
                  E-mail: husm@sdu.edu.cn        }


\begin{abstract}
In this paper, a total of about 28000 images in $V$ and $R$ band obtained on 161 nights using the one-meter optical telescope at Weihai Observatory (WHO) of Shandong University since 2008 to 2012 have been processed to measure the sky brightness. It provides us with an unprecedented database, which can be used to study the variation of the sky brightness with the sky position, the moonlight contribution, and the twilight sky brightness. The darkest sky brightness is about 19.0 and 18.6 $mag$ $arcsec^{-2}$ in $V$ and $R$ band, respectively. An obvious darkening trend is found at the first half of the night at WHO, and the variation rate is much larger in summer than that in other seasons. The sky brightness variation depends more on the azimuth than on the altitude of the telescope pointing for WHO. Our results indicate that the sky brightness at WHO is seriously influenced by the urban light.
\end{abstract}
\keywords{Astronomical Phenomena and Seeing}
\section{Introduction}
\label{sect:intro}

Sky brightness together with the number of clear nights, transparency, seeing, photometric stability, and humidity are the most important parameters used to describe the quality of an astronomical site \citep {patat03}. Sky brightness constrains the limiting magnitude of a telescope and the exposure time for the expected signal to noise ratio of the targets, and analysis of sky brightness in images made in the observatory can help to understand sources of noise well.  The study of the distribution of sky brightness and its variation with different factors due to light pollution is very important, not only in choosing proper object for observations, but also in estimating the appropriate exposure time for the given signal to noise ratio of the target. For these purposes, we have processed the data obtained in $V$ and $R$ band to measure the sky brightness by an automatic procedure.

Sky brightness is generated by several sources, such as astronomical twilight (namely airglow), scattering of starlight, zodiacal light and artificial light pollution, so it is a function of many variables, such as the altitude of the observation site, time on the time-scale of years as well as hours \citep{hampf11}.  Many previous works have been done to study the sky brightness, for example: \cite{walker88, krisciunas90, benn98, sanc07, neugent10,aceituno11, yao13, pedani14}. However, Weihai Oservatory is different from these observatories, since it is located at Weihai Torch Hi-Tech Science Park, which is only a few kilometers from the main city. So the most dominative light pollution source is the artificial light besides the moonlight when the moon is above the horizontal. Here we present $V$ and $R$ band sky brightness measurements for WHO, obtained on 161 nights from 2008 to 2012. These data provide us an unprecedented chance to investigate the relationship between sky brightness and the average airmass, azimuth, altitude, universal time, moon-target angular distance, moon elevation, moon phase and sun zenith distance.

The structure of this article is as follows. In Section 2, we give some information on the basic data reduction procedure; in Section 3, the photometric calibration and sky brightness calculation are described, while the sky brightness variations caused by different factors are described in Section 4. In Section 5, we summarize the results obtained in this work.

\section{Observations and data reduction}
\label{sect:data}
The one-meter Cassegrain telescope at Weihai Observatory was equipped with a back-illuminated PIXIS 2048B CCD camera from the Princeton Instruments Inc. and a standard Johnson/Cousins set of $UBVRI$ filters controlled by  a dual layer filter wheel from American Astronomical Consultants and Equipment. The PIXIS camera has $2048\times2048$ square pixels, and the pixel size is 13.5$\mu$m. The scale of the image is about 0.35$\arcsec$ per pixel, and the field of view is about 11.8$'$ $\times$ 11.8$'$. For more detailed information about WHO, see \cite{hu14}.

The data set discussed in this work have been obtained from February 2008 to December 2012, which are taken on 161 nights, using the one-meter telescope for accurate photometric purpose. During these five years, about 28000 images taken in $V$ and $R$ band were processed for sky brightness measurements. We usually tracked one target as long as possible during observations, so the sky brightness usually changed continuously if they were derived from the same day. Only these data for slow readout and low-noise output setup were used, the readout noise and gain under the same setup are 3.67 electrons and 1.65 electrons/ADU, respectively.  All frames were bias subtracted and flat-field corrected using standard IRAF\nolinebreak\footnotemark[1] procedures.\footnotetext[1]{IRAF is distributed by the National Optical Astronomy Observatories, which is operated by the Association of Universities for Research in Astronomy Inc., under contract to the National Science Foundation.} Since the dark current is virtually absent in the CCD detector with typical 30-300s integration times, no such correction was applied.

\section{Sky brightness calculation}
\label{sect:dataanalyse}
The procedure adopted to derive the sky brightness is based on simple photometric concepts.
After bias and flat correction, all the images were analyzed by the procedure. Firstly, it performed an automatic identification of the stars in each image using Naval Observatory Merged Astrometric Dataset (NOMAD) catalogs. Then WCS information was added to the FITs header by Automatic Stellar Coordinate fitting Package (ASCFIT) \citep{Jorgensen02}. All the stars in the same frame can be considered to have the same airmass, azimuth and altitude due to the small field view of our photometric system. So this procedure allows both to identify the stars and to estimate the corresponding airmass within the considered image. After that, the flux corresponding to each of these stars, whose signal to noise ratio are higher than 50, was derived by a simple aperture photometry algorithm using  DAOPHOT package in IRAF. The procedure provided us with a catalogue of detected stars, and the catalogue included the coordinates of the star projected on the sky, its airmass $\chi$, its catalogue magnitude $M$  at the corresponding band, and the instrumental  magnitude
$m = -2.5\;log_{10}(\frac{counts}{t_{exp}})$. These catalogues for one night can be used to derive the instrumental
zero-point ($ZP$) and the extinction coefficient ($\kappa$) for the
considered band ($x$), applying the following classical
formula with suitable airmass ranges:
 \begin{equation}
 M-m = ZP_{x} - \kappa_{x}  \chi
\end{equation}

\cite{harris81} recommend to use data spanning at least 1 airmass, but we had to relax this constraint just as \cite{patat03} did, because the altitude of our observations is usually higher than 30 degrees.
When we got all the values of ($M - m $) and $\chi$ for one night, a classical linear square-regression
between ($M - m $) and
the airmass ($\chi$) was performed. To eliminate clearly deviating data (such as variable stars), we have rejected all data deviating more than 1$\sigma$ (root mean square, RMS) and performed the least squares fitting again on the remaining data.  Fig. 1 shows an example of the least squares fitting.
\begin{figure*}
\includegraphics[angle=0,scale=0.58]{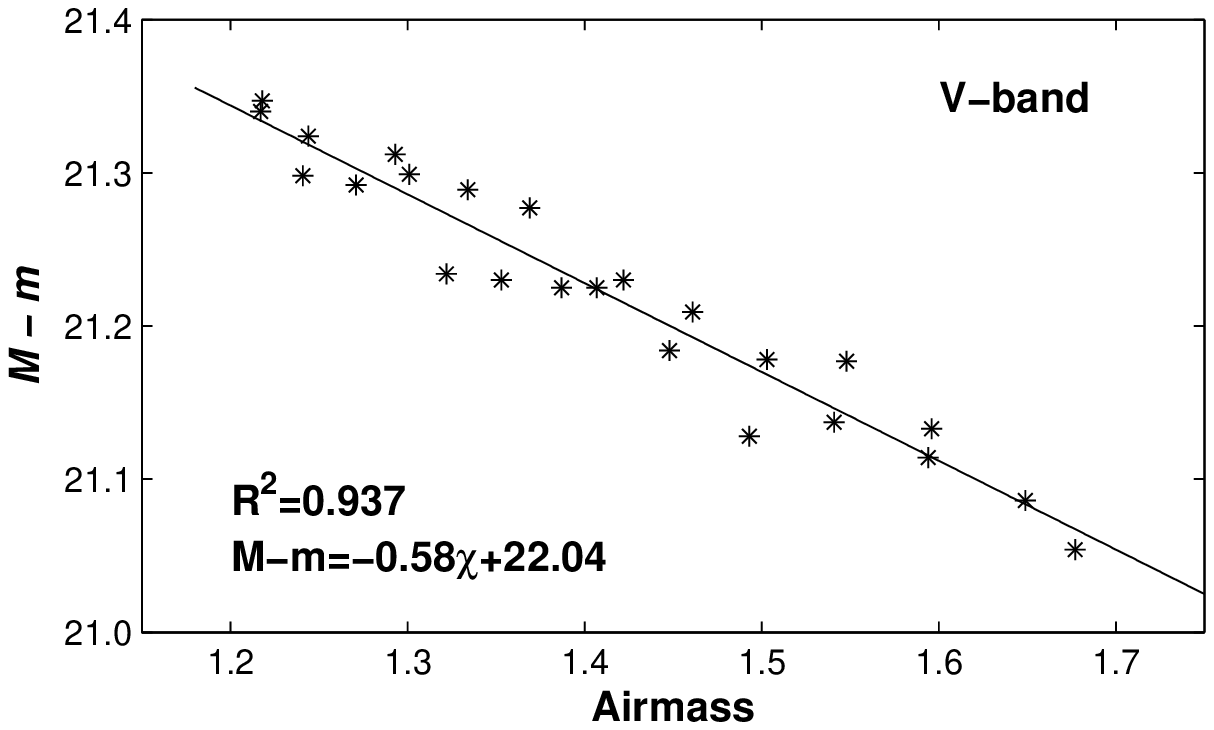}
\includegraphics[angle=0,scale=0.58]{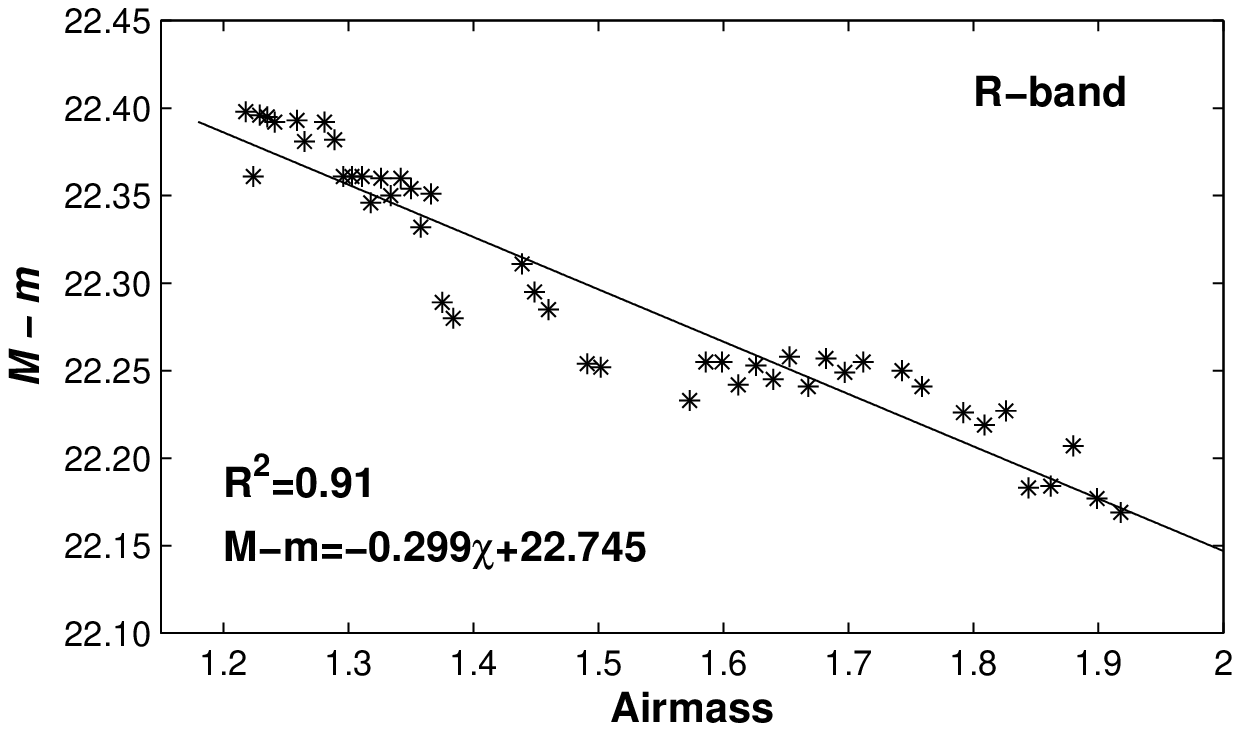}
\hfill \caption{Example of the linear regression technique used to derive the
zero-point at a certain night for considered filter. The difference between catalogue magnitude and instrumental magnitude versus the corresponding airmass is shown in the left and right panel for $V$ and $R$ band, respectively. }
\label{fit1}
\end{figure*}

The data were fitted by a linear regression
based on Equation (1). If the correlation coefficient of this regression is larger than 0.95, and the airmass range is larger than 0.4 as well, the zero-point of the corresponding filter is automatically updated. If this is not the case, adjacent zero-point can be used safely because the variations of zero-point are mainly due to the aging of telescope reflective surfaces and they are smaller than the inherent sky variations \citep {patat03}.

The instrumental sky brightness can be derived by the following formula:
\begin{equation}
\label{eq:skyb}
m_{sky} = -2.5\;log(\frac{I_{sky}}{scale^2 \; t_{exp}})
\end{equation}
 where $I_{sky}$ is the sky flux per pixel. The median sky values which help to reject stars and cosmic rays in the sky boxes, were used as $I_{sky}$ .
$Scale$ is the detector's scale (arcsec pix$^{-1}$), i.e. 0.35$\arcsec$ per pixel for our configuration,  and $t_{exp}$ is
the exposure time (in seconds).
The next step is to convert the instrumental magnitudes to the standard photometric
system. The sky brightness is calibrated without correcting the
atmospheric extinction, following the convention adopted in most recent studies of sky
brightness \citep[e.g.][]{walker88, krisciunas90, benn98,patat03,sanc07, neugent10,aceituno11, yao13, pedani14}.
Therefore, the calibrated sky brightness $M_{sky}$, can be given by
\begin{equation}
\label{equation2}
M_{sky} =ZP_{x} + m_{sky}
\end{equation}
where $ZP_{x}$ is the instrument zero point for a certain band derived by equation (1).

\section{Sky brighness distribution and variations}
Different factors can affect the sky brightness besides
city lights,  such as astronomical twilight, scattering of starlight, zodiacal light, the presence of atmospheric dust, the solar cycle, airmass, the phase and angular distance of the moon from the observed object, the Galactic and ecliptic latitude of the observation and the altitude and geomagnetic latitude of the observing site \citep{Taylor04}. However, with the rapid development of economy, human activity has added an extra source, namely the artificial light, to the sky brightness, especially for the observatories which are located nearby the city such as WHO. WHO is located at the campus of Shandong Universality (Weihai), and serious affected by the artificial light from campus and the main city of Weihai. So it is difficult to analyze the effect of zodiacal light, the solar cycle and so on. We mainly focus on how the moon and the artificial light affect the sky brightness variations.

To allow for a thorough analysis of the data, sky brightness measurements have been logged together with a large set of parameters, some of which are related to the target's position and others related to the ambient conditions. They include the average airmass, azimuth, altitude, universal time, sun zenith distance, moon phase, moon altitude and moon-target angular distance.

The histogram and cumulative statistic of sky brightness for $V$ and $R$ band using all the data for which sun zenith distance $\zeta > 106^{\circ} $(see Sec. 4.1) are shown in the left and right panel of Fig. 2. The median sky brightness for V and R band are 17.4 and 17.2 $mag$ $arcsec^{-2}$ , and the darkest sky brightness for $V$ and $R$ band can reach to 19 and 18.6 $mag$ $arcsec^{-2}$ , respectively. Fig. 2 shows that the sky brightness is darker than 17 $mag$ $arcsec^{-2}$ on about 70\% observations in both $V$ and $R$ band.

\begin{figure*}
\includegraphics[angle=0,scale=0.43]{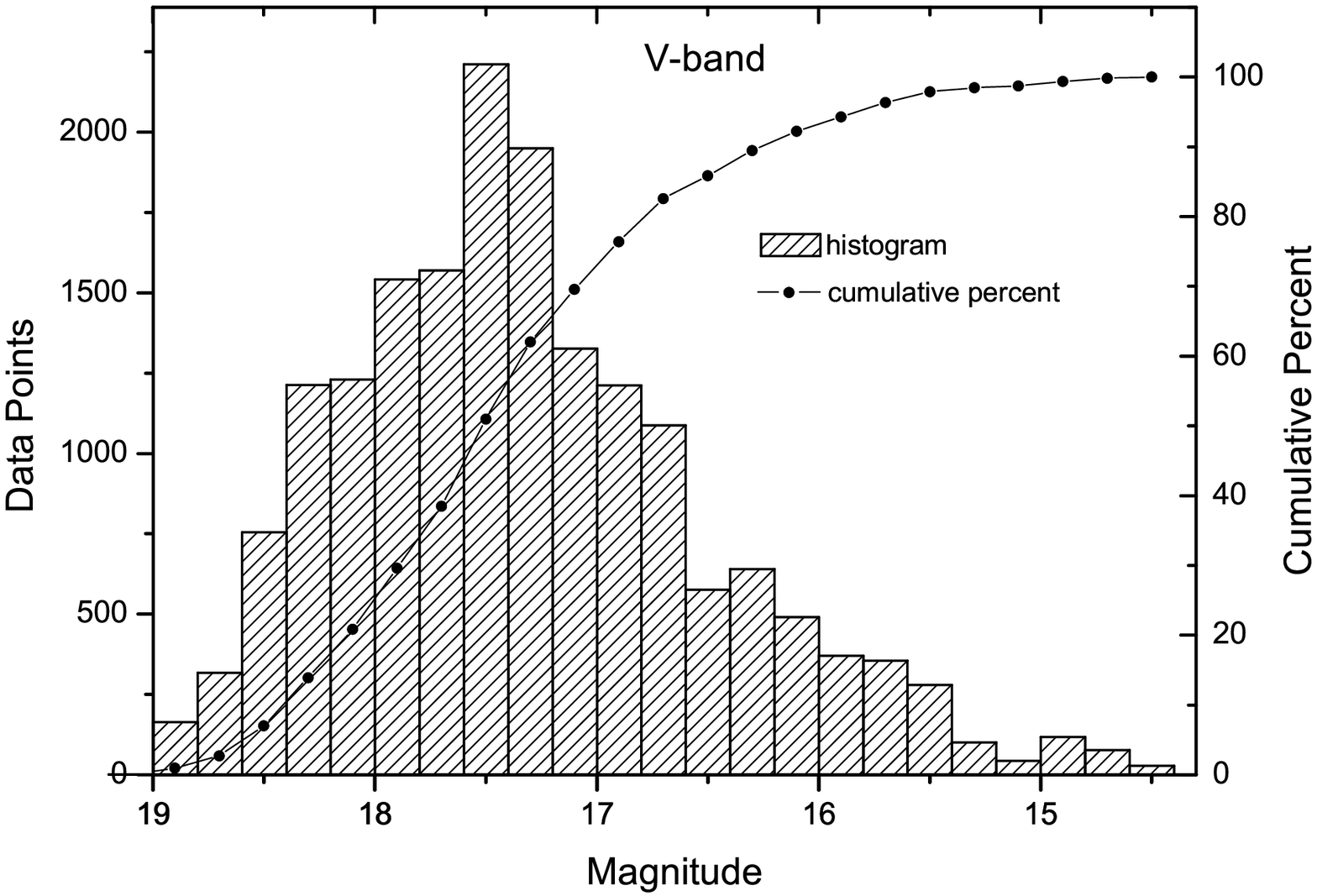}
\includegraphics[angle=0,scale=0.43]{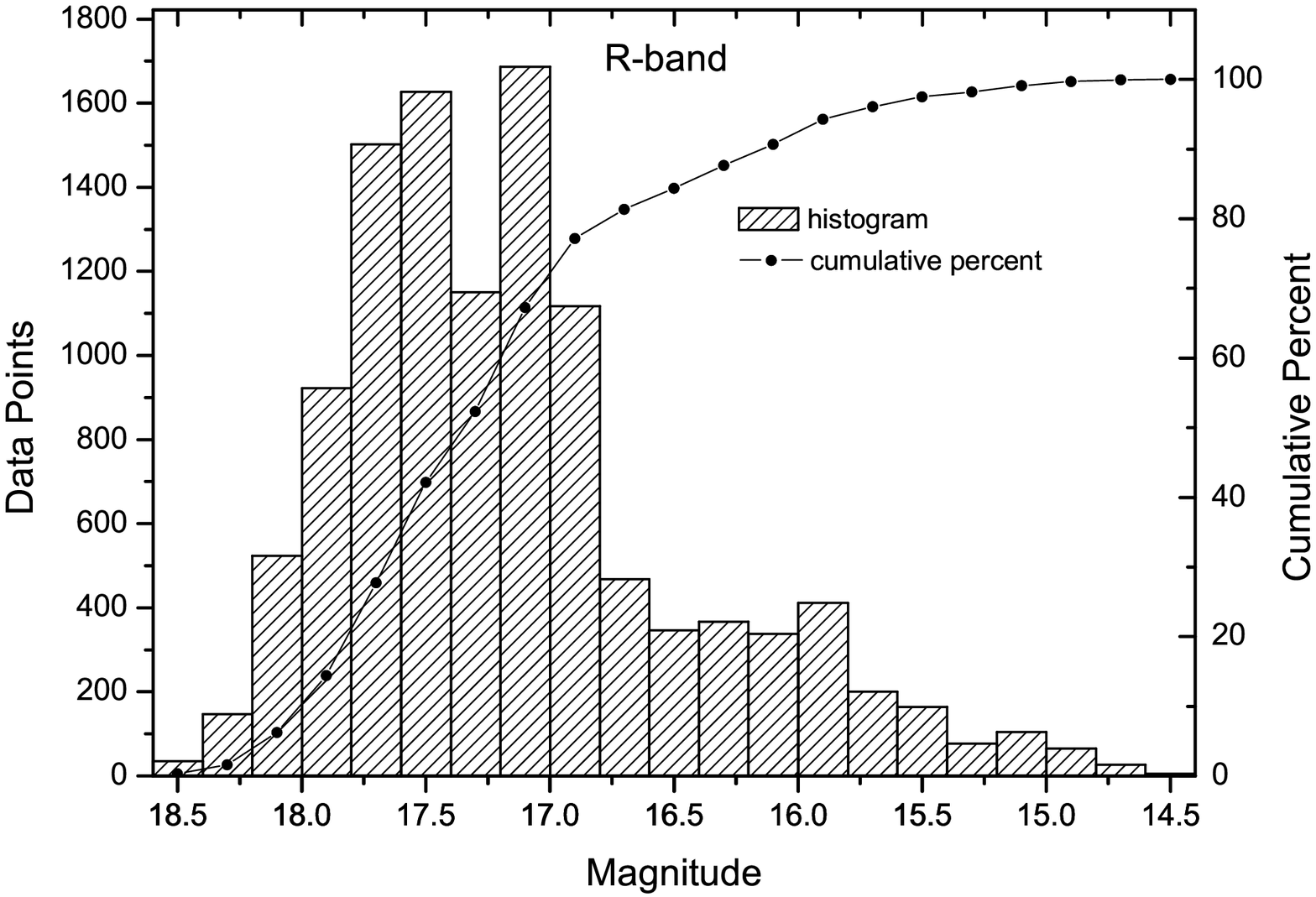}
\hfill \caption{Statistic results of sky brightness using all the data for which sun zenith distance $\zeta > 106^{\circ} $. Left panel is for $V$ band, and right panel is for $R$ band. Both panels show that the sky brightness is darker than 17 $mag$ $arcsec^{-2}$  for about 70\% observations.}
\label{fit1}
\end{figure*}

\subsection{Twilight sky brightness}
Since some data have been observed during the twilight, this offers us the possibility to measure its effect directly. To avoid contamination from scattered moonlight, we have chosen only those data points for which the moon was below the horizon. Since the twilight sky brightness changes with the position on the sky for a given sun zenith distance, it is necessary to make a selection on the data to study its behavior as a function of the sun zenith distance $\zeta$. In order to get sufficient measurements, we have used only these data points whose zenith distance satisfies $|\alpha|\leq40^{\circ}$. The results are presented in Fig. 3 for $V$ and $R$ band. One can see that the night sky brightness level is reached at around $\zeta=104^{\circ}-106^{\circ} $ in both of $V$ and $R$ band. That is to say the contribution of sun light can be safely neglected when the sun altitude is lower than $-$16 degrees.

In order to give a more quantitative description of the observations, we did a linear fitting to the twilight data for $\zeta \leq 101^{\circ} $,  when the contribution by the night sky is still moderate to get the slope $\gamma$. The slope $\gamma$ is 1.22 $mag$ $deg^{-1}$ for $V$ band, and 1.13 $mag$ $deg^{-1}$ for $R$ band. To convert the values of $\gamma$ into brightness variation per unit time, we use the formula given by \cite{patat06}
\begin{equation}\label{}
\frac{\mathrm{d}\varphi}{\mathrm{d}t}=\cos\varphi\sin{H_{\odot}}\cos\delta_{\odot}\cos\phi\frac{\mathrm{d}H_{\odot}}{\mathrm{d}t},
\end{equation}
where $\varphi=\zeta-90$,  $\phi$ is the site latitude, $H_{\odot}$ and $\delta_{\odot}$ are the hour angle and the declination of the sun, respectively. We can obtain ${\mathrm{d}\varphi}/{\mathrm{d}t}= 0.20$ deg $min^{-1}$ at the equinoxes ($\delta_{\odot}=0$) for WHO ($\phi=37.5^{\circ}$) for $\varphi\sim0$ (i.e. at the time of sunrise and sunset), where ${\mathrm{d}H_{\odot}}/{\mathrm{d}t}\simeq0.25$ deg $min^{-1}$. Applying this factor to the slope values, one can obtain brightness decline rates for $V$ and $R$ band which are 0.24 and 0.23 $mag$ $arcsec^{-2}$ $min^{-1}$, respectively. These values can provide us with useful information to compute an ideal sequence of exposure times for twilight skyflats in future \citep{tyson93}.

\begin{figure*}
\includegraphics[angle=0,scale=0.65]{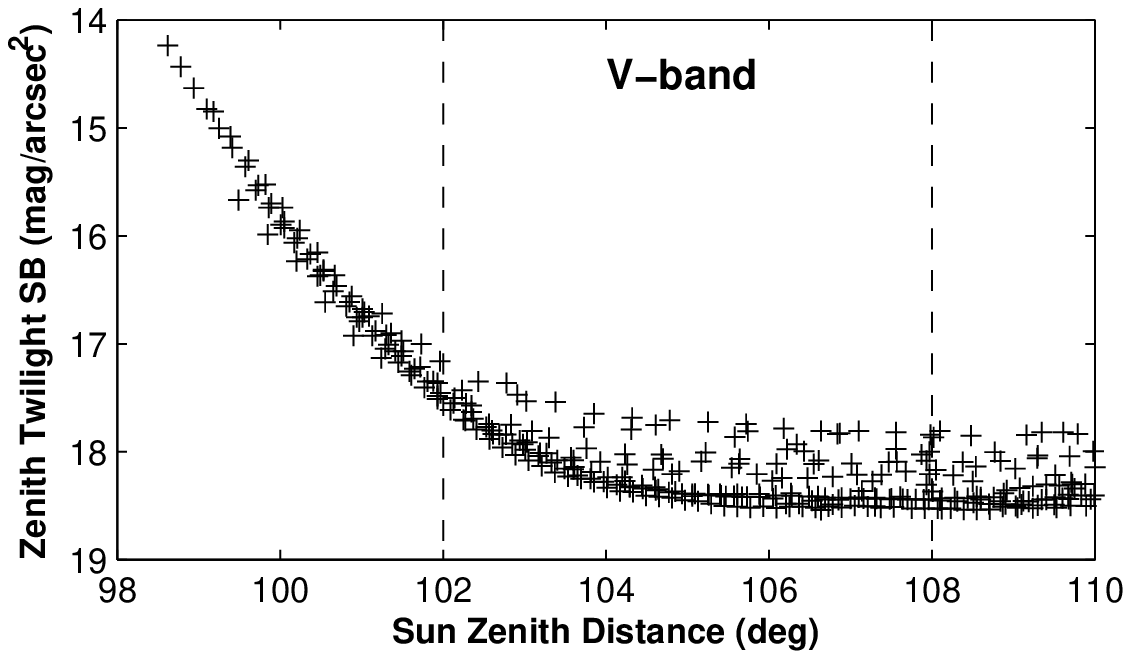}
\includegraphics[angle=0,scale=0.65]{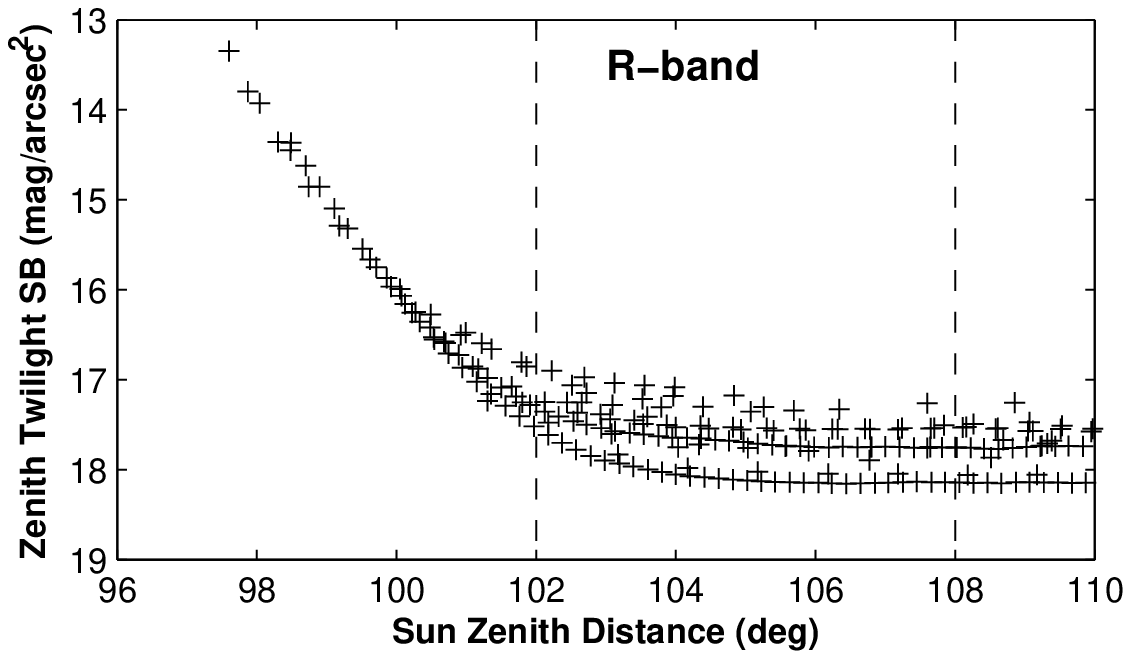}

\hfill \caption{Zenith twilight sky brightness (SB) for $V$ and $R$ band in the left and right panel, respectively. The vertical dashed lines mark the end of nautical ($left$) twilight and astronomical ($right$) twilight.}
\label{fit1}
\end{figure*}

\subsection{Moon contribution}
Another relevant aspect is the contribution produced by scattered moonlight when measuring the sky brightness. The effect of the moon on the sky brightness is a complicated function of the phase of the moon, the altitude of the moon, the angular distance between the moon and the target, and the atmospheric extinction \citep{krisciunas91}. Data that have been collected when the moon contribution to the sky brightness is conspicuous, offer us the possibility to measure its effect directly.
To investigate the effect of scattering light from the moon, the sky brightness derived from images with sun zenith distance $\zeta > 106^{\circ} $, moon phase $>$ 0.5 (1 is for full moon), moon altitude$ > 5^{\circ}$ and UT $>$ 16 (After the middle night, the artificial light almost remained constant, and it can minimize the effect caused by the artificial light, see Sec. 4.4) is plotted against moon-target angular distance in the left panel of Fig. 4. Although there are some fluctuations, we can still see an obvious trend that the sky brightness is becoming fainter with enlarging separation angular distance. The sky brightness reaches its minimum when  the separation angular distance is at around 90 degrees. After that the sky brightness seems to be constant. The shape of sky brightness versus the moon-target separation angular distance is just like the scatter function $f(\rho)=f_{R}(\rho)+f_{M}(\rho)$, where $f_{R}(\rho)$ and $f_{M}(\rho)$ are the Rayleigh and Mie scattering functions, respectively. $\rho$ is the scattering angle defined as the angular separation between the moon and the sky position \citep{krisciunas91}.
Deviations of 1 $mag$ can be seen even at the same angular distance and at roughly the same moon phase (see the right panel of Fig. 4). So it is clearly not enough to predict the sky brightness with sufficient accuracy using Walker's function \citep{walker87}, since it has only one input parameter, namely the moon phase. The sky brightness depends on a number of parameters, some of which  are known only when the target is going to be observed, such as the local extinction coefficient, the angular separation of the moon, sky position and so on. So the model proposed by \cite{krisciunas91} is much more promising, since it takes all relevant astronomical circumstances into consideration.
Meanwhile, we also analyze the sky brightness variations versus lunar altitude using all the data for $V$ and $R$ band. The results are displayed in Fig. 5. It shows that the sky brightness in $V$ and $R$ band are nearly constant for lunar altitudes smaller than 0 degree, i.e. below the horizon, and then brighten quickly until the lunar altitude ascends to about 40 degrees. After that, the sky brightness changes little. Namely, when the moon altitude reaches 40 degrees, the sky brightness seems to reach constant on the whole.

  \begin{figure*}
\includegraphics[angle=0,scale=0.60]{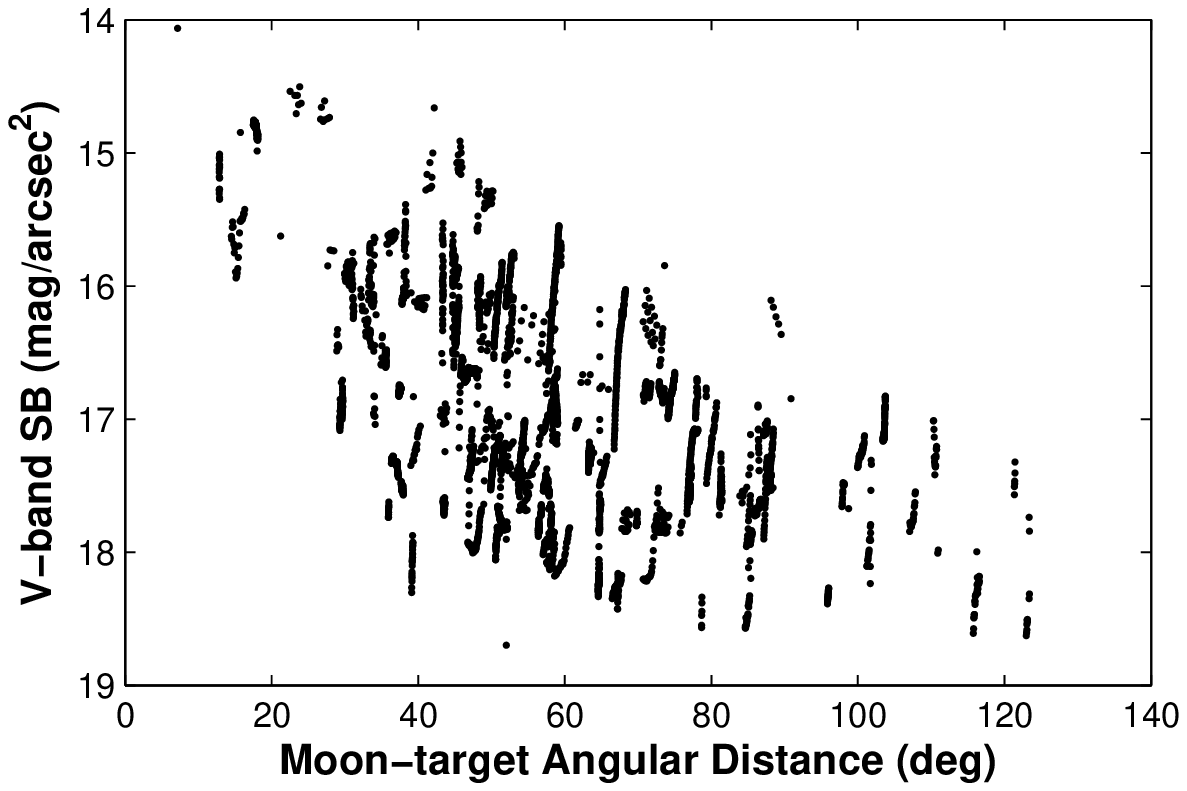}
\includegraphics[angle=0,scale=0.60]{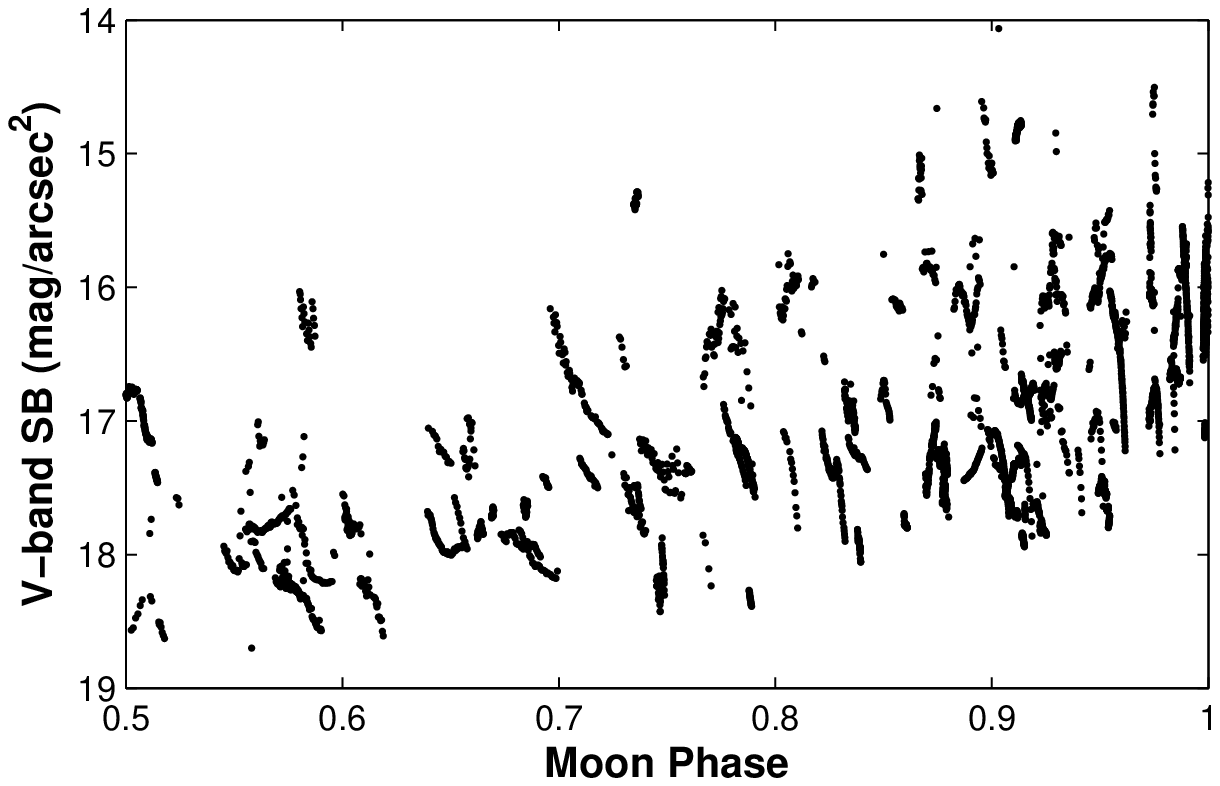}

\hfill \caption{Left panel: the distribution of the sky brightness for
$V$ band as a function of the moon-target angular distance. Right panel: sky brightness versus moon phase using the same data.}
\label{fit1}
\end{figure*}

  \begin{figure*}
\includegraphics[angle=0,scale=0.65]{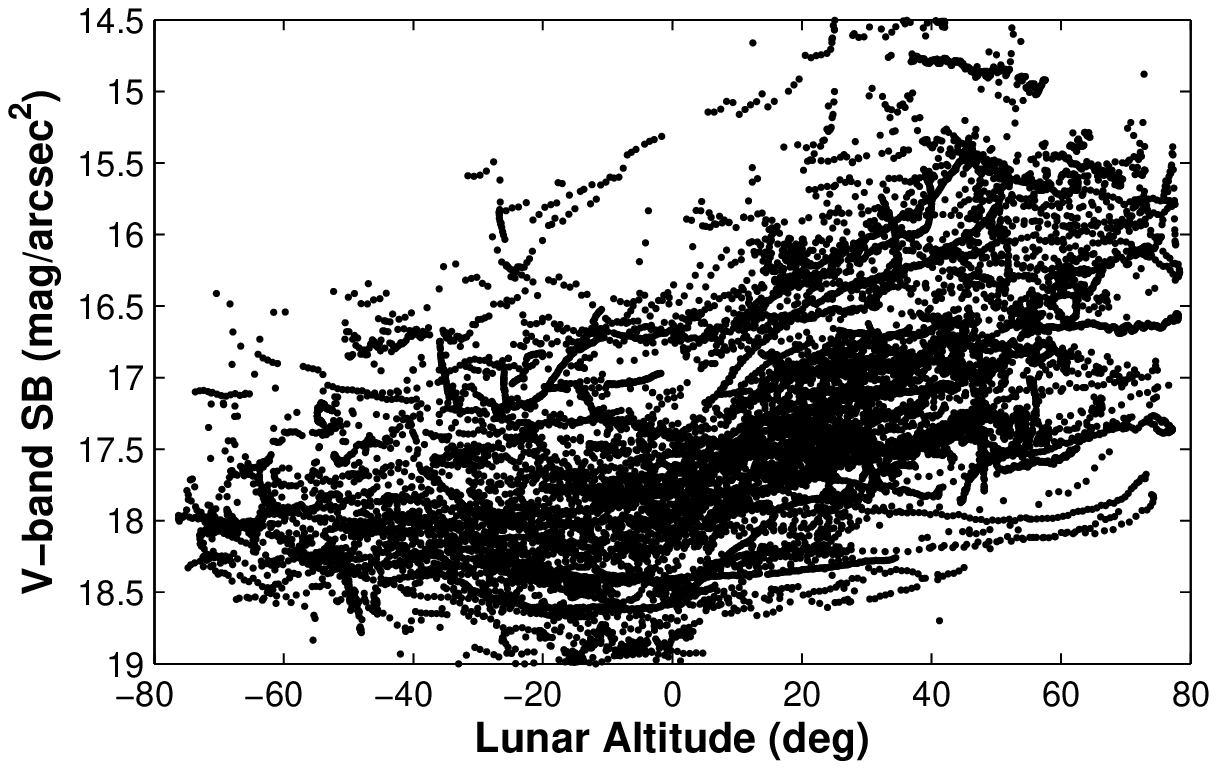}
\includegraphics[angle=0,scale=0.65]{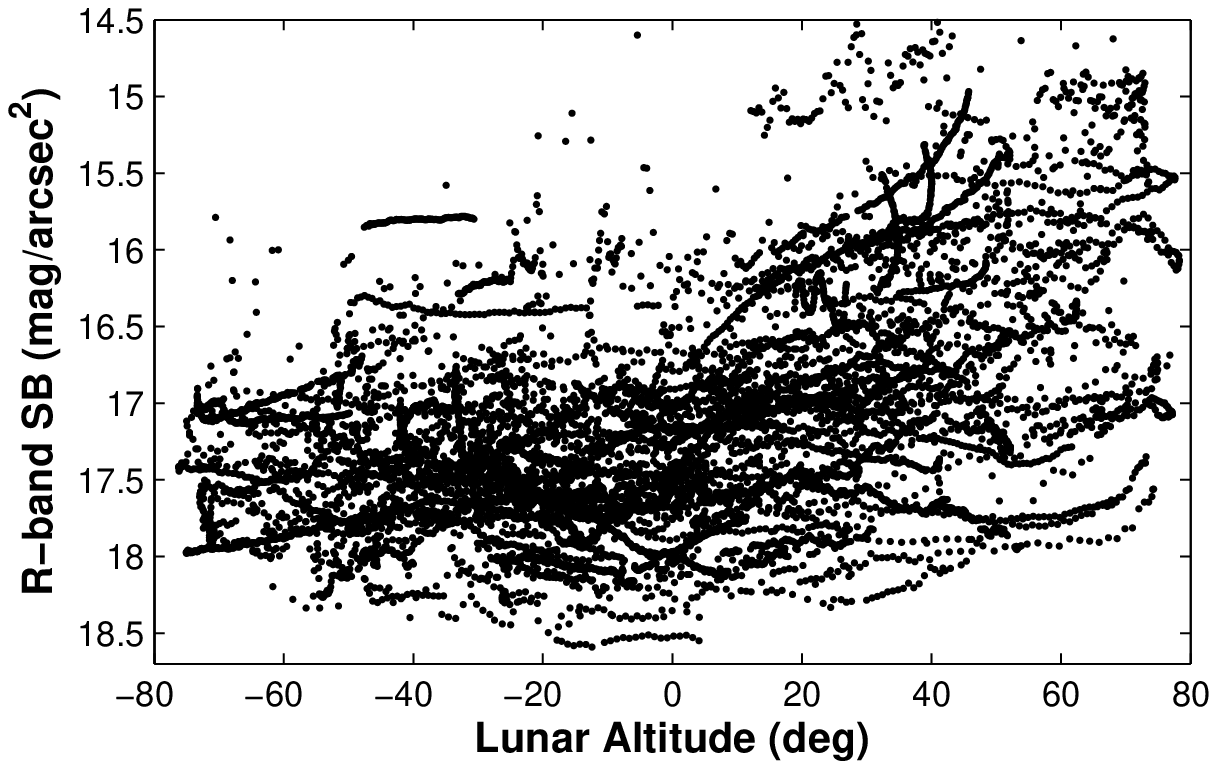}
\hfill \caption{Variations of sky brightness versus the lunar altitude. Left panel use the data derived from $V$ band and right panel use the data derived from $R$ band. }
\label{fit1}
\end{figure*}

\subsection{Sky brightness variations versus altitude and azimuth}
In order to investigate the sky brightness variation versus the altitude of the telescope pointing during the dark time, it is necessary to apply some selections on the data since they were observed under a wide range of conditions. For this purpose, we have adopted the following criteria:  the sun zenith distance $\zeta$ is larger than $106^{\circ} $ and the moon altitude is below the horizon. The results obtained from this selection are shown in Fig. 6. As can be seen in this plot, the sky brightness remains nearly constant when the altitude becomes higher than 40 degrees for most data set. This phenomenon indicates that the altitude has little impact on the sky brightness for WHO.

To determine how city lights affect the sky brightness, the sky brightness versus azimuth of the observations is shown in Fig. 7. In order to minimize the effect of altitude, only those data whose altitudes are higher than 40 degrees is used. Azimuth coordinates are measured in degrees from the south point westwards (i.e. west is at $90^{\circ}$, north at $180^{\circ}$, east at $270^{\circ}$). Fig. 7 shows that images observed toward north direction (azimuth around from 110 to 230 degrees) tend to have darker skies than that in other directions.  The sea is just in north direction, so the skies are less affected by the artificial light compared with other directions. The sky brightness begins to brighten from  azimuth 200 to 250 degrees, and darken from 250 to 300 degrees. It reflects that the light pollution is very serious in this direction, and the main city (Huancui District) is just located in this direction.

The sky brightness strongly correlated with azimuth of the telescope pointing at WHO compared with altitude, and it suggests that the surrounding lighting environments of the observing sites have significant effects on the observation.

  \begin{figure*}
\includegraphics[angle=0,scale=0.65]{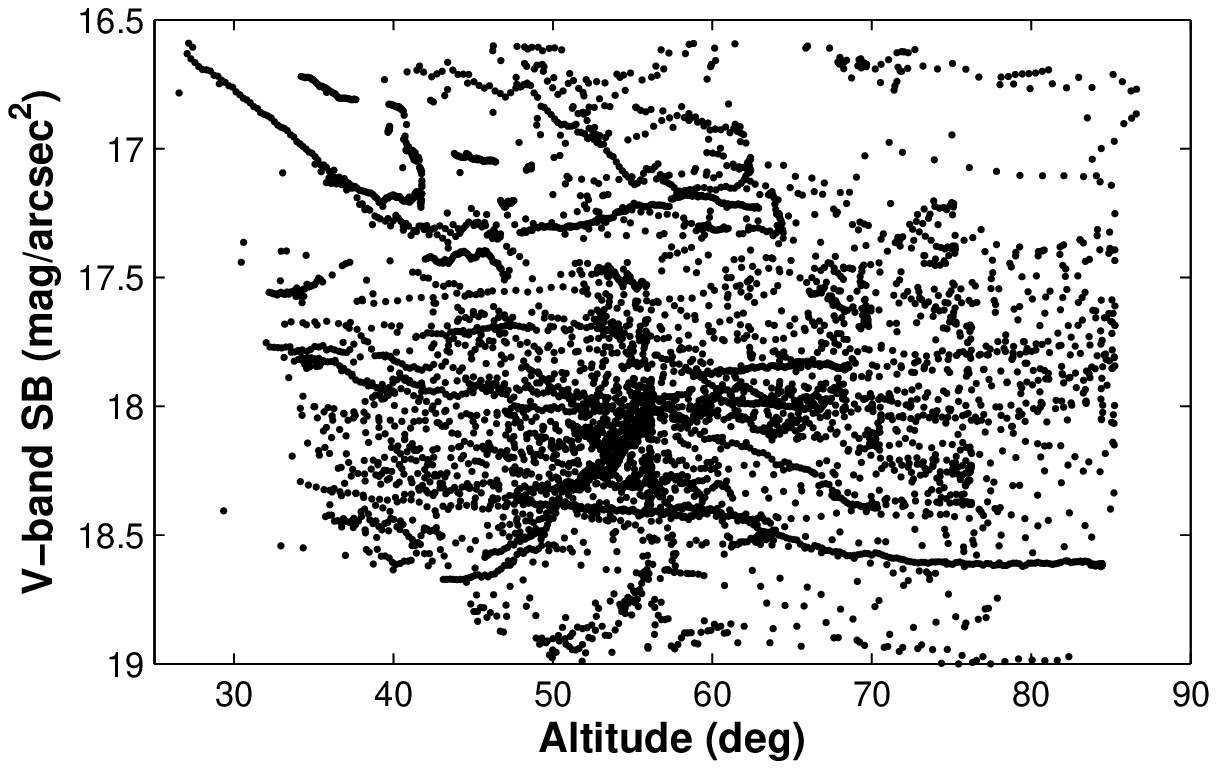}
\includegraphics[angle=0,scale=0.65]{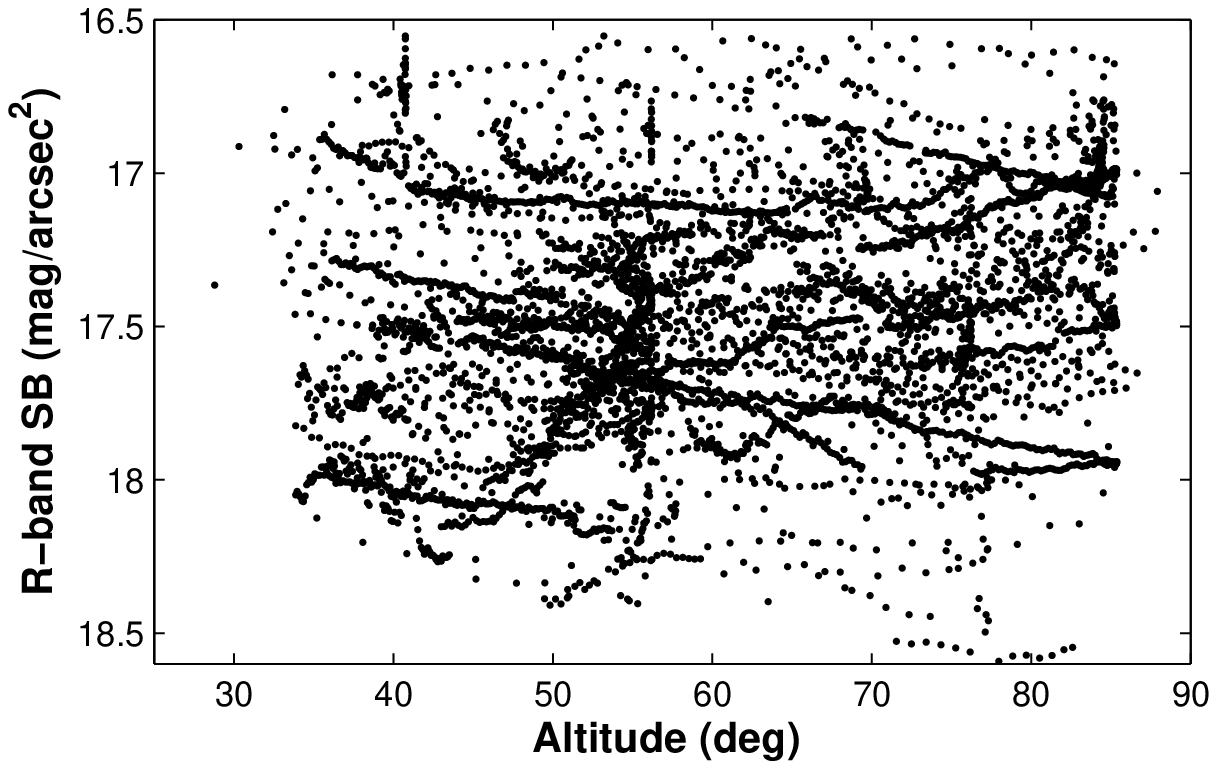}
\hfill \caption{Distribution of the sky brightness as a function of altitude of the telescope pointing
for $V$ and $R$ band in the left and right panel, respectively.}
\label{fit1}
\end{figure*}

 \begin{figure*}
\includegraphics[angle=0,scale=0.65]{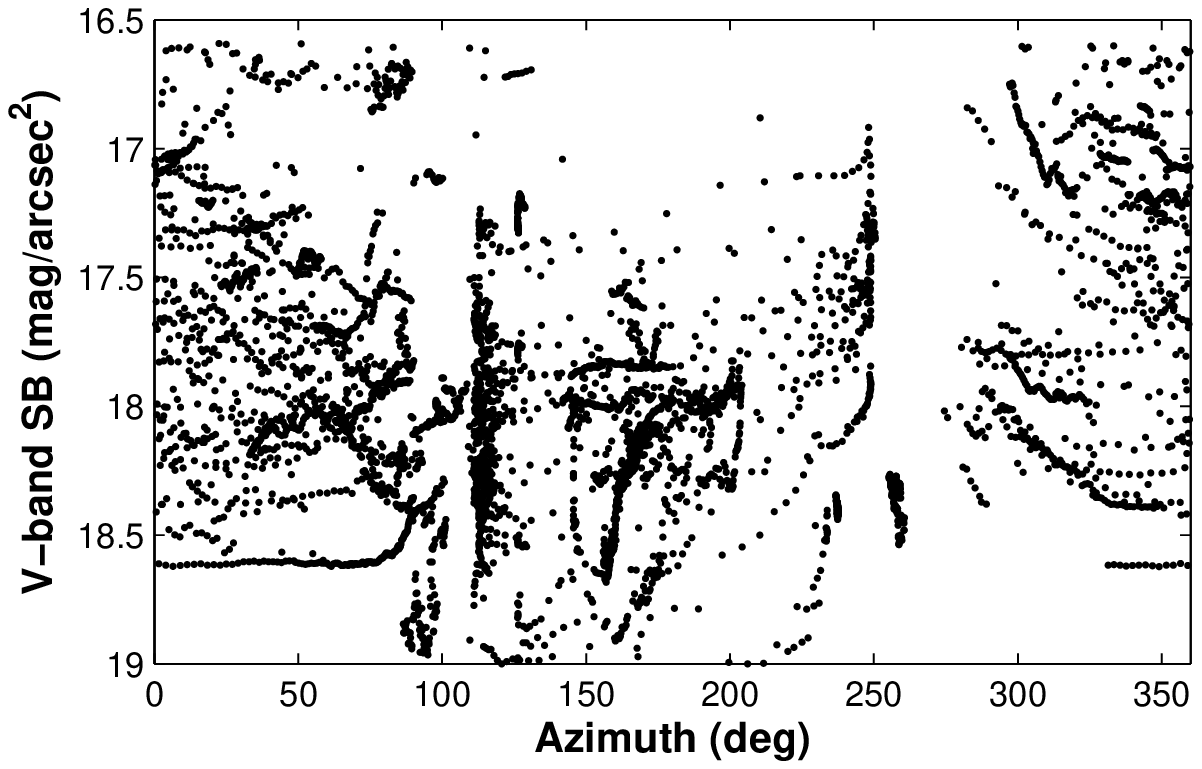}
\includegraphics[angle=0,scale=0.65]{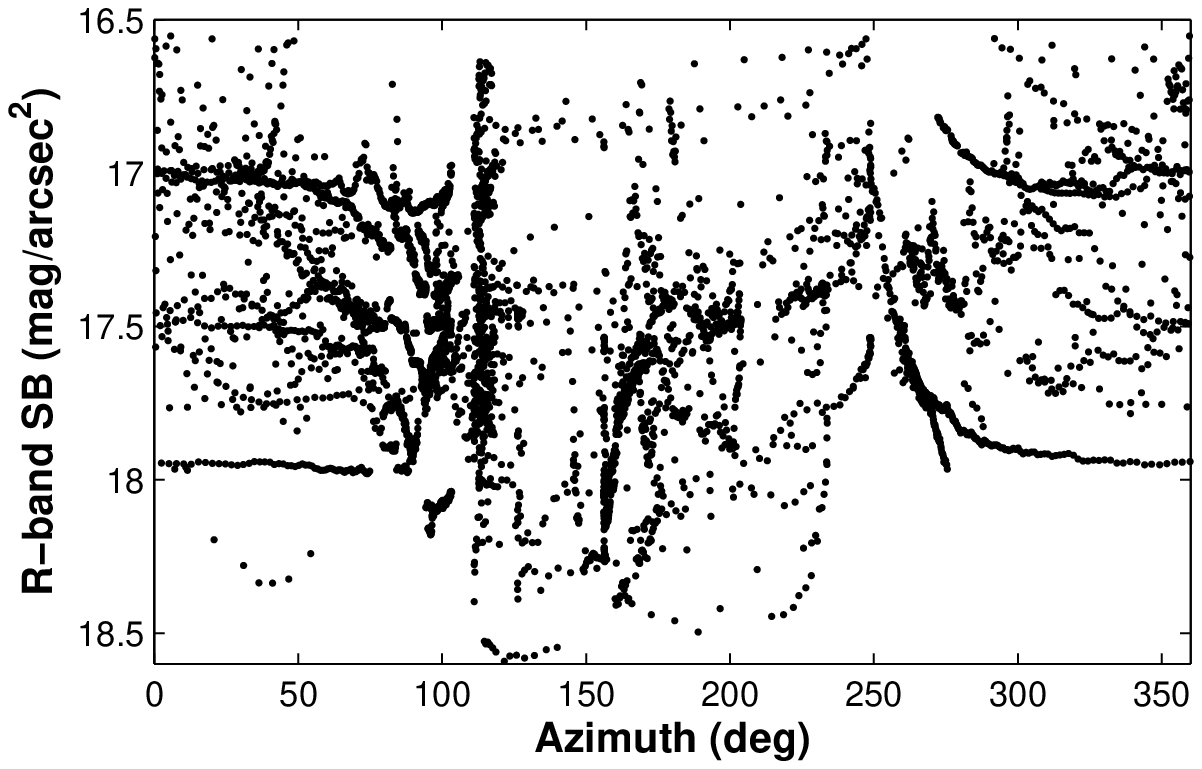}

\hfill \caption{Distribution of the sky brightness versus azimuth of the telescope pointing
for $V$ and $R$ band in the left and right panel, respectively.}
\label{fit1}
\end{figure*}

\subsection{Sky brightness variations during the night}
 \begin{figure*}
\includegraphics[angle=0,scale=0.65]{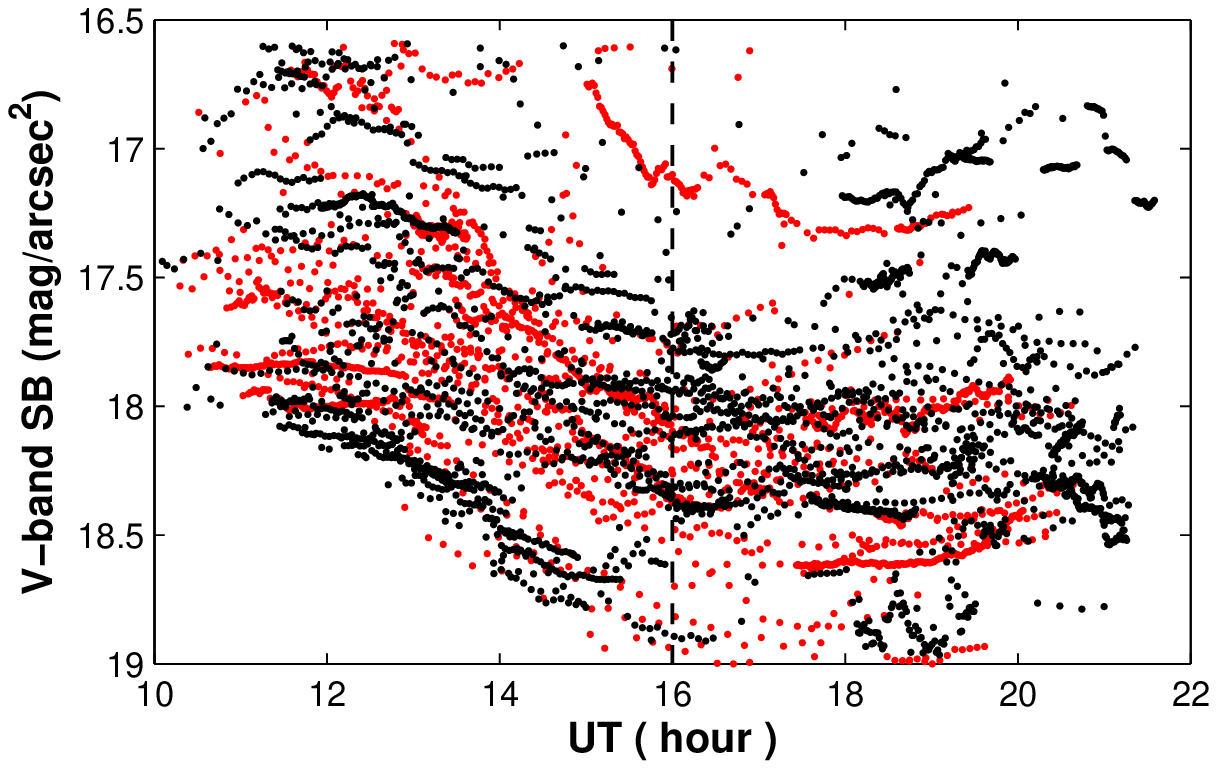}
\includegraphics[angle=0,scale=0.65]{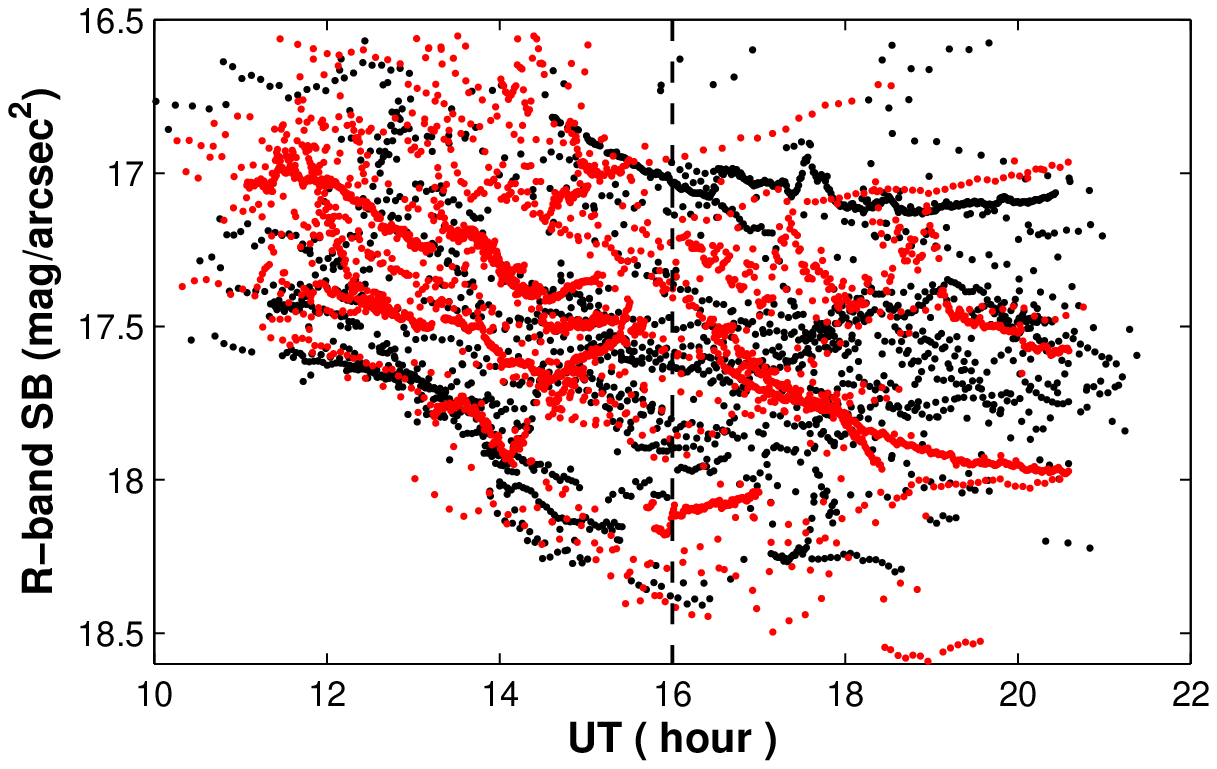}
\hfill \caption{Sky brightness as a function of universal time for $V$ and $R$ band in the left and right panel, respectively. The red dots represent data derived from summer (ranging from May to October). The black dots represent the rest data. The vertical dashed lines mark the midnight for local time.  }
\label{fit1}
\end{figure*}
Many discordant results about the sky brightness variations as a function of the time distance from astronomical twilight have been presented in the past by different authors at different observing sites. \cite{walker88} found a steady decrease of $\sim0.4$ $mag$ $arcsec^{-2}$ during the first six hours after the end of twilight, and concluded that it is more likely due to a natural phenomenon than a decrease in the contribution of city lights throughout the night. \cite{krisciunas90} found that the $V$ band data showed a decrease of $\sim0.3$ $mag$ $arcsec^{-2}$ during the first six hours after the end of twilight, but he also remarked that it was not clearly seen in the $B$ band.
Due to the high artificial light pollution, \cite{lockwood90}
tend to attribute the nightly sky brightness decline they observed at the Lowell Observatory, to continuous reduction of commercial and private activities. \cite{Taylor04} also found a decrease of sky brightness in the first half of the night in $V$ and $R$ band. However, \cite{leinert95} and \cite{mattila96} did not find any significant evidence of decreasing sky brightness after the end of twilight. More recently, \cite{patat03} also did not find any evidence at least in $V$ and $R$ band, and he pointed out that findings by \cite{walker88} were probably influenced by a small number of well sample nights. \cite{pedani09} did not find any
significant correlation between the sky brightness and
the fraction of the night, and concluded that the results derived by \cite{Taylor04} could be in part due to the lower statistics.

In order to investigate the sky brightness variation during nights,
we have chosen only those data points for which the moon is below the horizon and the sun zenith distance $\zeta $ is larger than $106^{\circ} $, avoiding contamination from scattered moonlight and the twilight. At the same time, the altitudes are higher than 40 degrees. The selected measurements are shown in Fig. 8. As one know, Weihai is a well-known harbor and tourist city, in summer seasons (from May to October) lots of people come to Weihai for vacation and more artificial outdoor lightings work on at that time, so we use red dots (derived from May to October) for comparison. From Fig. 8, one can see that the night sky brightness has an obvious slowly darkening tendency from the beginning of observation, then it seems to be constant after about 16:00 UT ( Beijing time is 24:00). Due to the fact that our data set collects observations performed under a wide range of conditions, such as different azimuth, altitude of target and weather conditions, the sky brightness deviation is very large even at the same universal time. The slope of the linear trend is 0.138$\pm0.01$ $mag$ $hr^{-1}$  and 0.106$\pm0.017$ $mag$ $hr^{-1}$ for $V$ and $R$ band using the data ranging from UT 13:00 to 16:00 (In order to give uniform weighting to different time, we only choose the data ranging from 13 to 16 O'clock. Because winter is darkening earlier than summer), respectively.  The gradual trend of darkening of sky brightness before the midnight corresponds primarily to the continuous reduction of commercial and private activities when the night wears on. As businesses close and some of their signs are turned off, traffic declines, and people go to bed, the artificial light becomes fainter and fainter. On the other hand, street lights and some signs are illuminated all night. After midnight, the artificial light remains constant, so the sky brightness also reaches constant at the same time on the whole. This tendency strongly supports that the nightly sky brightness decline at WHO is associated with human activities, which is similar to the results derived by \cite{lockwood90} at the Lowell Observatory.

From the left panel of Fig. 8, we see that the sky brightness in $V$ band has a steeper slope in summer comparing with other seasons (the remainder of the year). The results of the slope for the linear trend are 0.23$\pm0.01$ $mag$ $hr^{-1}$  and 0.09$\pm0.02$ $mag$ $hr^{-1}$ in $V$ band for summer and other seasons, respectively. This phenomenon indicates that more public and private light sources are switching on in summer due to human's outdoor activities. From the right panel of Fig. 8, one can see that the sky brightness in $R$ band has a different behavior, having much more fluctuations before the midnight comparing with other seasons. This is probably caused by different lighting patterns of the external public or private light source.

\section{Results and discussions}
\label{sect:results}
In this work, we systemically present measurements of the sky brightness at Weihai Observatory  based on a large data set collected from the accurate photometric observations from 2008 to 2012.  The night sky brightness can reach as faint as 19 $mag$ $arcsec^{-2}$  and 18.6 $mag$ $arcsec^{-2}$  under best conditions for $V$ and $R$ band, respectively.

The following results were derived by this work. Firstly, A steady decrease of $\sim1.0$ $mag$ $arcsec^{-2}$  is found during the first six hours before the midnight for $V$ band and $\sim0.7$ $mag$ $arcsec^{-2}$ for $R$ band. The variation rate is 0.138 $mag$ $hr^{-1}$  and 0.106 $mag$ $hr^{-1}$ for $V$ and $R$ band, respectively. The slope of the linear trend is more steeper in summer than that in other seasons. The gradual trend of darkening of sky brightness before the midnight correspond primarily to the continuous reduction of commercial and private activities as the night wears on. With some public and commercial lightings are turned off, traffic declines, and people go to bed, the sky brightness reaches constant. These results show that human activities have become a major problem for our observation.
Secondly, the sky brightness is different at different directions. In the sea side the sky brightness is much darker. Finally, the faintest sky brightness is only about 19 $mag$ $arcsec^{-2}$  and 18.6 $mag$ $arcsec^{-2}$  for $V$ and $R$ band, respectively.  These phenomena can be explained by the influence of human activities on the observation. The altitude and the angular distance of the moon also play an important role in the sky brightness besides the moon phase. Twilight influence on the sky brightness can be neglected when the sun zenith distance $\zeta$ is larger than $106^{\circ}$.
\section*{Acknowledgments}
 We acknowledge the anonymous referee for his/her comments and suggestions that lead to a better manuscript. We also thank Dr. Y. G. Jiang and K. Li for their kindly English editing. This work is partly supported by the National Natural Science Foundation of China under grants No. 11203016, 11333002 and by the National Natural Science Foundation of China and Chinese Academic of Sciences joint fund on astronomy under project No. 10778701, 10778619.  This work is partly supported by the Natural Science Foundation of Shandong Province under grant No. ZR2012AQ008.

\label{lastpage}

\end{document}